\documentclass[
 aip,
 amsmath,amssymb,
 reprint,%
]{revtex4-2}

\usepackage{graphicx}
\usepackage{dcolumn}
\usepackage{bm}
\usepackage{float}
\usepackage[utf8]{inputenc}
\usepackage[T1]{fontenc}
\usepackage{mathptmx}
\usepackage{etoolbox}
\usepackage{amsmath}
\usepackage{hyperref}
\makeatletter
\def\@email#1#2{%
 \endgroup
 \patchcmd{\titleblock@produce}
  {\frontmatter@RRAPformat}
  {\frontmatter@RRAPformat{\produce@RRAP{*#1\href{mailto:#2}{#2}}}\frontmatter@RRAPformat}
  {}{}
}%
\makeatother
\begin{document}

\preprint{AIP/123-QED}
\title{ Evidence of orbital Hall current induced correlation in second harmonic response of longitudinal and transverse voltage in light metal-ferromagnet bilayers }
\author{Dhananjaya Mahapatra}
 \altaffiliation{Department of Physical Sciences,
 Indian Institute of Science Education and Research Kolkata, Mohanpur, West Bengal,741246, INDIA.}
 \author{Harekrishna Bhunia}
 \altaffiliation{Department of Physical Sciences,
 Indian Institute of Science Education and Research Kolkata, Mohanpur, West Bengal,741246, INDIA.}
\author{Abu Bakkar Miah}
 \altaffiliation{Department of Physical Sciences,
 Indian Institute of Science Education and Research Kolkata, Mohanpur, West Bengal,741246, INDIA.}

  \author{Soumik Aon}
 \altaffiliation{Department of Physical Sciences,
 Indian Institute of Science Education and Research Kolkata, Mohanpur, West Bengal,741246, INDIA.}
\author{Partha Mitra}
 \email{pmitra@iiserkol.ac.in}
\altaffiliation{Department of Physical Sciences,
 Indian Institute of Science Education and Research Kolkata, Mohanpur, West Bengal,741246, INDIA.}

\date{\today}
\begin{abstract}
We investigate the effect of orbital current arising from orbital Hall effect in thin films of Light Metals of  Nb and Ti in ohmic contact with ferromagnetic Ni in the second harmonic longitudinal and transverse voltages, in response to an a.c. current applied to the bilayer structures. Analogous to the experiments on Heavy Metal-Ferromagnet bilayers, we extract the Orbital Hall Torque efficiency and 
 unidirectional orbital magnetoresistance (UOMR) per unit current density. Control experiments on light metal bilayers with NiFe (permalloy) as ferromagnet and a single layer of Ni shows absence torque and supression of the UOMR, corroborating the role of Ni as an efficient converter of orbital to spin current and role of Light metals as source of orbital current. We find that both Orbital Hall torque efficiency and unidirectional orbital magnetoresistance are proportionately  higher in Ti as compared to Nb , confirming a correlation between the two effects arising from the same Orbital current in the respective light metals.

\end{abstract}

\maketitle

\begin{figure}
\centering
    \includegraphics[width=.5\textwidth]{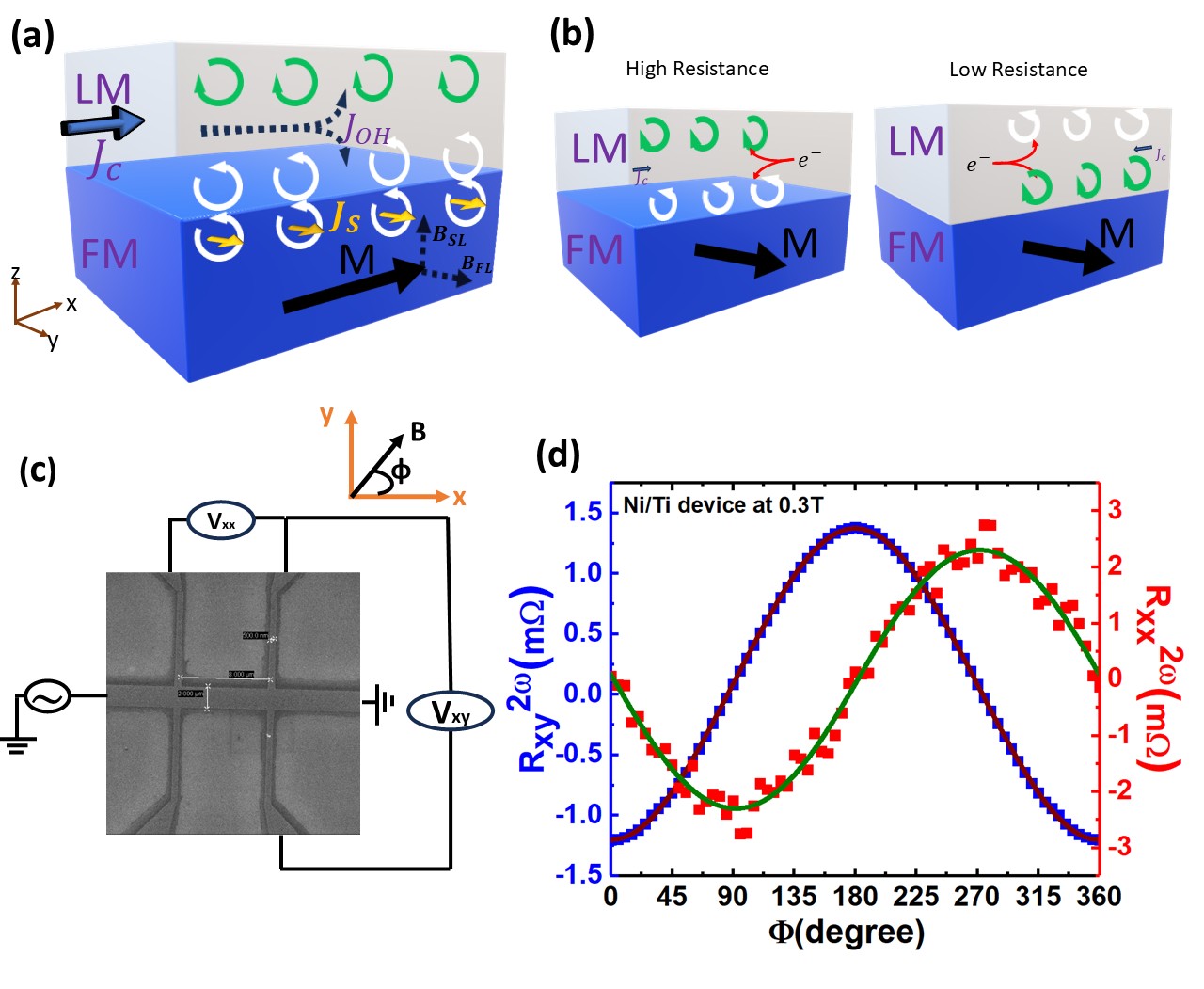}
    \caption{(a).Schematic representation illustrates the formation of the orbital Hall effect(OHE) and subsequent transfer of Orbital current to Spin current within the Ni layer. This process results in the application of torque on the magnetization. (b).Parallel alignment results in high resistance, while antiparallel alignment yields low resistance, demonstrating unidirectional magnetoresistance.  (c). Schematic of the measurement setup of the Bilayer Hallbar device. (d). Simultaneous measurements of transverse Hall and longitudinal second-harmonic signals were obtained using a lock-in amplifier of FM/LM bilayer device.  }
    \label{1}
\end{figure}

A current flowing through heavy metals (HM) with appreciable spin-orbit coupling (SOC) strength generates transverse pure spin current due to the phenomenon of spin Hall effect (SHE) \cite{PhysRevLett.83.1834, PhysRevLett.85.393, PhysRevLett.92.126603,6516040}, which is confirmed through numerous experiments\cite{15539563, JOUR04937, PhysRevLett.119.087203}. A standard experimental technique that manifests SHE induced spin currents is the measurement of second harmonic voltages in response to an a.c. current flowing through the plane of bilayer heterostuctures HM/FM, where FM is a layer of ferromagnetic conductor\cite{PhysRevB.90.224427}. The polarization axis of the spin current generated in the HM due to SHE is linked only to the direction of current flow which remains constant, while the magnetisation of the FM is rotated continuously with the help of an external magnetic field.
The second harmonic Hall voltage is shown to be linked to magnetisation dynamics in the FM induced by the transfer of the component of spin angular momentum from the injected spin currents arising in the adjacent HM layer, perpendicular to the magnetization of the FM. The effect of the component of the spin current along the magnetisation also reveals itself in the longitudinal voltage, which is referred to as the unidirectional spin Hall magnetoresistance (USMR)\cite{Avci2015, PhysRevLett.121.087207, 10.1063/1.4935497, 10.1063/1.4983784} and can be grossly understood as a spin valve type effect in the second harmonic response,  with resistance varying from minimum to maximum when the magnetisation is rotated from being parallel to antiparallel with respect to the spin current polarization. 
\\Recently theoretical works have pointed out the fact that the carriers in a conductor also possess orbital angular momentum (OAM) in addition to spin angular momentum and predict a new phenomenon analogous to SHE, where carriers with opposite OAM  are segregated in a direction transverse to current flow and is termed as orbital Hall effect (OHE)\cite{PhysRevLett.121.086602, PhysRevB.98.214405, PhysRevResearch.2.013177}. It is further emphasized that the OHE will be relevant even in conductors with insignificant SOC, classified as Light Metals (LM). Such predictions have prompted a series of experiments on LM/FM bilayer structures on the exact same lines as that of HM/FM bilayers mentioned previously\cite{Lee2021, Lee2021nature, PhysRevResearch.4.033037, PhysRevB.107.134423, Choi2023, Hayashi2023}. However, unlike the spin, the OAM of a carrier cannot directly interact with the magnetisation of the FM.  It is now established that Nickel as the FM layer exhibits special characteristics that can effectively convert the OAM current injected from the LM layer into a spin current\cite{PhysRevResearch.2.033401, Lee2021, Lee2021nature, PhysRevB.107.134423}. Although the exact underlying mechanism is yet to be established, it is speculated in some reports that this unique ability of Ni can be attributed to its large SOC compared to other popular FMs like Py, Co, CoFe etc. that form the basis of most reported works on HM/FM bilayers. Once the OAM current is converted to a spin current, the usual picture of a torque acting on the magnetisation of Ni is applicable and is termed as Orbital Torque due to its origin. Similarly, the resistance of LM/Ni bilayers exhibits a change in resistance as the magnetisation orientation is flipped laterally with respect to current flow and the phenomenon is termed as unidirectional orbital magnetoresistance (UOMR)\cite{PhysRevResearch.4.L032041}.
\\\\
Here we present a systematic investigation of OAM current on a set of Hall bar shaped FM/LM bilayers (Fig. \ref{1}(c)) where we have chosen a 3d element Titatnium (Ti) and a 4d element Niobium (Nb) as the LM, motivated by the recent theoretical calculations that show both Ti and Nb has almost zero spin Hall conductivity but a finite orbital Hall conductivity\cite{salemi2022first}, which is also confirmed by recent experiments\cite{Choi2023, PhysRevB.107.134423}. As FM layer we have used both  Ni and NiFe for a given LM for comparison with the motivation that devices with NiFe will not convert the OAM current to spin current and hence manifest the effect of OAM current in the form of spin torque. Further, we have conducted simultaneous measurement of second harmonic longitudinal and transverse resistance, which is expected to arise from the mutually perpendicular components of the incident orbital current with respect to the FM magnetization.  A typical variation of the second harmonic resistances for a Ni/Ti device from $B=0.3$T  shown in Fig. \ref{1}(d) reveals that the variation of $R_{xy}^{2\omega}$ is opposite to that of $R_{xx}^{2\omega}$. Following the standard analysis followed in the case of spin currents in FM/HM bilayers we extract the torque efficiency and $\Delta R_{UMR}$ per unit current and present a correlation between the two quantities.

\begin{figure}[h!]
    \includegraphics[width=.5\textwidth]{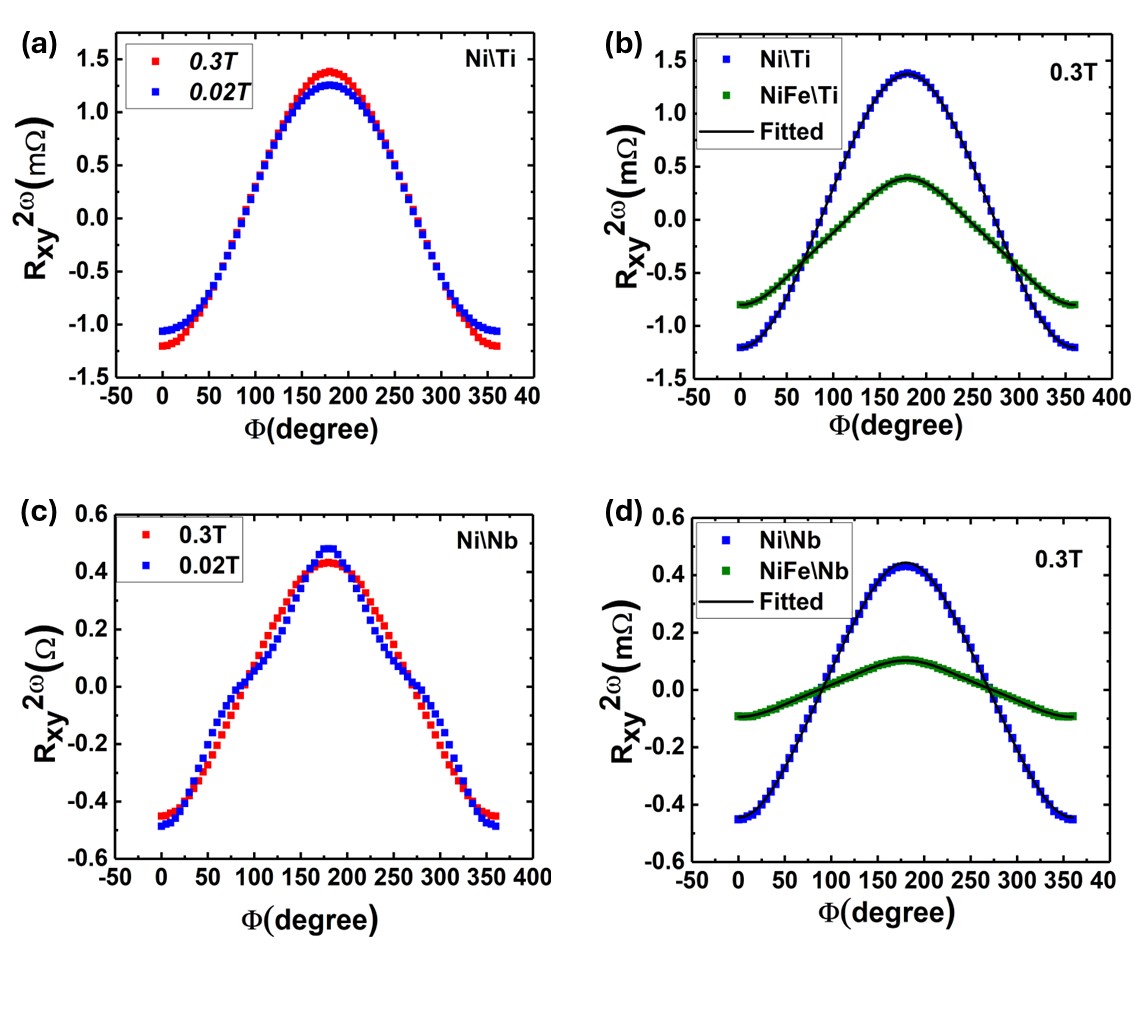}
    \caption{The angular dependence of second-harmonic Hall resistance ( of FM/LM bilayers during in-plane rotation of external magnetic field, strong enough to rotate the magnetisation of FM. Panels (a) and (c) show the angular variations at the two extreme magnetic fields for typical Ni/Ti and Ni/Nb devices, respectively. The relative magnitude of ($R^{2\omega}_{xy})$ in FM/LM bilayers using Ni and NiFe as FM  for Ti [panel (b)]  and Nb [panel (d)] as LM, respectively, for the highest magnetic field} 
    \label{2}
\end{figure}

The first harmonic transverse voltage response of the FM/HM bilayers is a linear effect that depends only on  the equilibrium state of magnetization.  The in-plane and out-of-plane components of the magnetization results in Planar Hall Effect(PHE) and Anomalous Hall Effect(AHE) respectively, which contributes to the $R_{xy}^\omega$ depending on the relative angle between magnetization, controlled by the external magnetic field ($B_{ext}$), with respect to current. The second harmonic voltage response arises from precessional motion of the magnetization around the equilibrium position due to toque induced by the spin current\cite{PhysRevB.90.224427,PhysRevB.100.214438} injected from the adjacent HM either due to bulk Spin Hall Effect or Rashba Edelstein type inetrface effects. As the second harmonic voltage is non-linear in current, the corresponding resistance $R_{xy}^{2\omega}$ is  dependent on current it is customary to calculate effective fields per unit current that exerts the spin orbit torques per unit magnetization, also known as torque efficiencies.
In FM/LM bilayers, the LM layer doesnot generate any spin current under an applied current bias due to lack of appreciable Spin-Orbit coupling,  Instead a transverse OAM current $J_{OH}$ is generated and injected into the adjacent FM layer\cite{PhysRevLett.121.086602}. However, unlike the spin currents ,  the orbital current does not directly interact with the  magnetization of the FM and hence will not reveal itself in the form of a torque under ordinary circumstances. Recent findings suggest that certain ferromagnetic materials e.g Ni \cite{Lee2021, Lee2021nature, Choi2023, PhysRevResearch.2.033401} and Co \cite{PhysRevResearch.4.L032041},  can facilitate the conversion of orbital current into spin current with certain efficiency, possibly due its inherent spin-orbit coupling (SOC).  This spin current will induce magnetization dynamics similar to the situation in FM/HM and the resulting torque is termed as Orbital torque and can be analyzed on the same footing as SOT\cite{PhysRevB.100.214438} and the angular dependence of $R_{xy}^{2\omega}$ for FM layers with  in-plane magnetization, is expressed by the phenomenological equation, by\cite{PhysRevB.99.195103, JOUR, AHN202312}
\begin{equation}
\begin{split}
    R_{xy}^{2\omega}(\Phi)=[R_{AHE}(\frac{B_{SL}}{B_{ext}+B_{eff}^k})+\alpha B_{ext} + R_{\nabla T}]cos(\Phi)\\
    +2R_{PHE}(\frac{B_{FL}+B_{Oe}}{B_{ext}})cos(2\Phi)cos(\Phi)
    \end{split}
    \label{eq1}
\end{equation}
where,$B_{SL}$ and $B_{FL}$ are the  Slonczewski like and Field like effective fields that exerts in-plane and out-of-plane torque  on the magnetization respectively.  $R_{AHE}$ is the anomalous Hall resistance, $B_{Oe}$ is the Oersted field , $B_{eff}^{k}$ is the effective anisotropy field of the FM layer. $\alpha$ is the ordinary Nernst co-efficient and $R_{{\nabla}T}$ is the anomalous Nernst  co-efficient, which are spurious heating effect terms due to large current density required in the experiments.

We have chosen Ti, a 3d element and Nb, a 4d element as LM material with possible different orbital textures. Each LM is paired separately with Ni and NiFe as FM layer to exploit the fact  that unlike Ni, NiFe  is not known to exhibit orbital to spin conversion. Hence, LM bilayers with NiFe will act as control samples which should not display any effect of OHE, while the samples with Ni should show evidence of Orbital Torque. Fig. \ref{2} (a) and (c) shows the unprocessed data of $R_{xy}^{2\omega}$ for Ni/Ti and Ni/Nb devices for two extreme values of $B_{ext}$ and in both cases we observe the characteristic features of magnetization dynamics described by Eq. \ref{eq1}, which in this case is attributed to Orbital Torque. In Fig. \ref{2}(b) and (d) we compare the unprocessed data of $R_{xy}^{2\omega}$ at $B_{ext}=0.3T$ for FM/Ti and FM/Nb bilayers respectively. The data although clearly shows reduction in the magnitude of $R_{xy}^{2\Omega}$ for the bilayers with NiFe as compared to those of Ni, but is still not zero as one would expect under ideal conditions.  This is due to the presence of significant heating effects which is unavoidable for these experiments. The effect of torque can be segregated from the heating effects by their characteristic dependence on $B_{ext}$, while obtaining the values of $R_{PHE}$ from $R_{xy}^\omega(\Phi)$ and $R_{AHE}$  and $B^k_{eff}$ from standard AHE measurement with out-of-plane magnetic field ($B_z$) in the range $\pm1T$ (see supplementary note 2).
\begin{figure}[h!]
    \includegraphics[width=.5\textwidth]{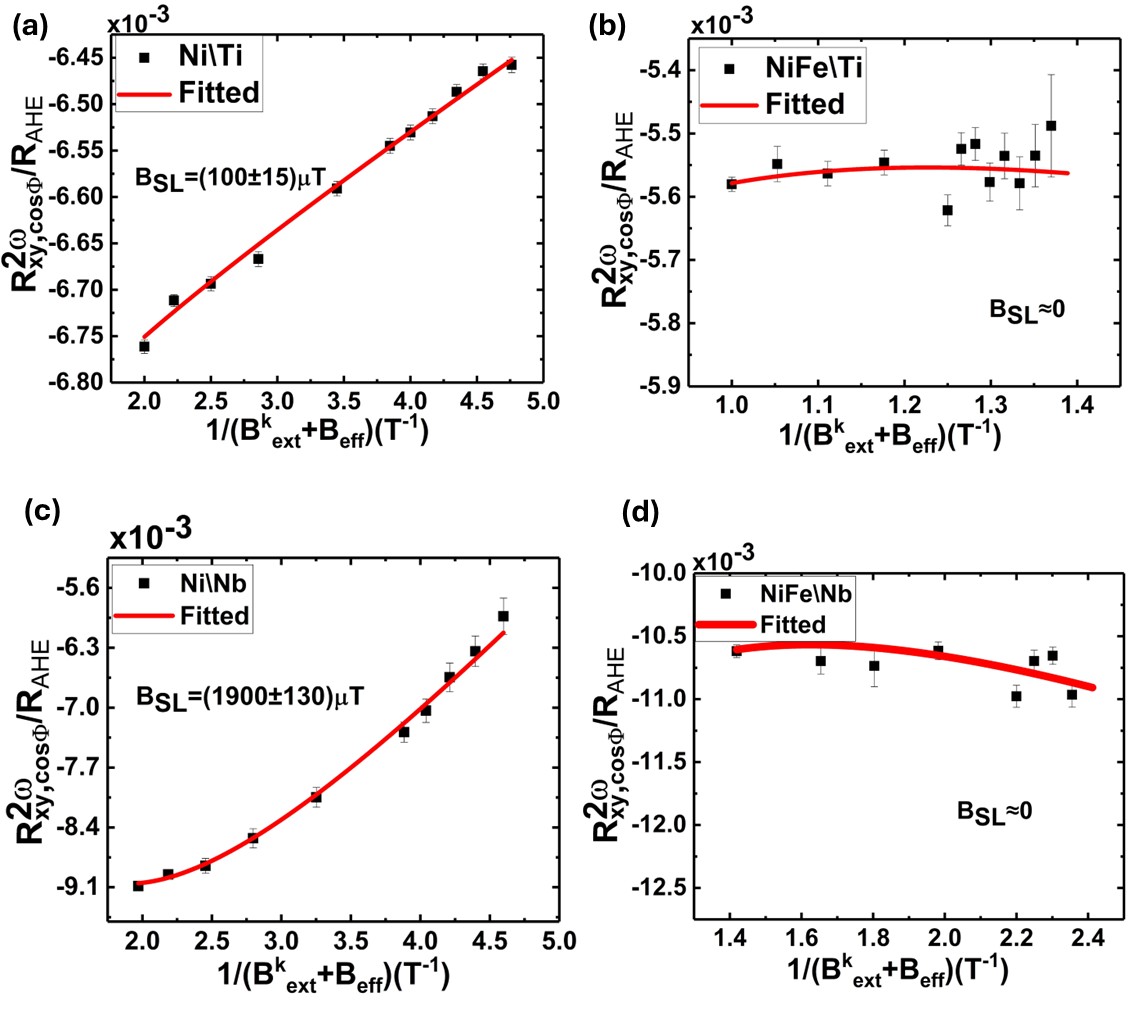}
    \caption{Plot of coefficients of obtained by fitting  Eq. (1) to the angular variation scans of the second-harmonic Hall resistance for fixed $B_{ext}$ as a function of for the FM/LM devices as indicated. $R_{AHE}$ and $B^k_{eff}$ are extracted separately from anomalous Hall effect measurements using same applied currents. The dominant nonlinear behaviour observed typically in  Ni/Nb devices [panel (c)]  is indicative of a large Ordinary Nernst Effect (ONE) directly proportional to in. Panels (b) and (c) confirm the absence of torques in NiFe/Ti and NiFe/Nb, respectively.}
    \label{3}
\end{figure}
Fig. \ref{3} shows a plot of the coefficient of $ cos(\Phi)$ with $\frac{1}{B_{ext}+B^k_{eff}}$ which reveals the true character of orbital torque and the fits of the data following Eq.\ref{eq1}  leads to an estimate of $B_{SL}$. A pronounced dependence on $B_{ext}$ observed in bilayers with Ni (Fig. \ref{3} (a) ,(c)) further corroborates the idea that Ni efficiently converts the resulting OAM current into a spin current leading to an observable orbital torque and $B_{SL}=100\pm15\mu$T and $1900\pm130\mu$T for Ni/Ti and Ni/Nb respectively. On the contrary the data for the NiFe bilayer samples donot show any systematic dependence on $B_{ext}$ indicative of the absence of torque effect, which can be understood to be due to the inability of NiFe to convert the OAM current into spin current. To check weather the results in Ni/LM bilayers are not a property of the Ni layer only, we also analyzed  the second harmonic Hall resistance data on a single layer of Ni and found that the equivalent orbital torque is very small, $B_{SL}=3\pm1\mu$.  To compare the orbital torque on different  devices which may be bearing varied current densities, it is customary to calculated the torque efficiency per unit applied current density as\cite{PhysRevB.92.064426}

\begin{equation}
    \zeta_{SL}=\frac{2e}{\hbar} M_s t_{\small{FM}} \frac{B_{SL}}{J_{LM}}
\end{equation}
We find $\zeta_{SL}$ for Ni/Ti and Ni/Nb  and Ni single layer to be $0.04\pm0.006$,$0.018\pm0.001$ and $(6\pm1)10^{-4}$respectively. This further establishes the fact that the LM is the source of the orbital Torque acting on the Ni layer in Ni/LM bilayers and not a property of only Ni and that Ti has a higher OHE conductivity than Nb. We are not emphasizing on the analysis of FL torque which is embedded in the  $cos(2\Phi)cos(\Phi)$ term of Eq. \ref{eq1} as it is dominated by interface properties of the FM/LM rather than bulk properties of LM like OHE conductivity\cite{PhysRevB.92.064426, PhysRevApplied.13.054014} ( See Supplementary note 4 ).

In the context of spin transport in FM/HM bilayer devices, a characteristic $\hat{m}\times\hat{j}$ dependence in the second harmonic response  of longitudinal resistance $R_{xx}^{2\omega}$ is termed as Unidirectional Spin Hall Magnetoresistance (USMR) \cite{Avci2015, PhysRevLett.121.087207, 10.1063/1.4935497, 10.1063/1.4983784}. The analogous phenomenon in a FM/LM system was recently reported \cite{PhysRevResearch.4.L032041} in Cu/Co bilayers and termed as Unidirectional Orbital Magnetoresistance (UOMR).  The underlying mechanism of USMR (or UOMR) is the presence of accumulated spin (or OAM) at the interface of FM with the normal metal, HM(or LM), due to which the overall resistance of the bilayer varies depending on the relative orientation of the FM magnetization with  respect to current direction, similar to CIP GMR phenomenon in FM/NM/FM trilayer devices.  The  resistance of the bilayer is maximum when the majority spins of the ferromagnet is aligned parallel to the accumulated moment, spin (or OAM (Fig. \ref{1}a)) and minimum in the antiparallel configuration (Fig. \ref{1}b).  However, there are additional contribution to the $R_{xx}^{2\omega}$, and the overall angular dependence is found to obey the  expression\cite{PhysRevB.106.094422, PhysRevLett.127.207206}
\begin{equation}
    R_{xx}^{2\omega}(\Phi)=R^* sin(\Phi)-2\Delta R_{xx}^{1\omega} \frac{B_{FL}+B_{Oe}}{B_{ext}} cos^2(\Phi) sin(\Phi)
    \label{equation3}
\end{equation}
The coefficient of $sin(\Phi)(\sim m_y)$ includes the effect of UMR ( either USMR or UOMR) and thermal gradient, $R^*=gR_{\nabla T} + R_{UMR}$. Experimentally, the thermal gradient contribution $R_{\nabla T}$ is independently extracted from the analysis of $R^{2\omega}_{xy}$  and scaled by the  geometric factor  $g = l/w$ , $l$ and $w$ being the length and width of the hall bars respectively. In our devices, $g\approx4$ which is also consistent with the ratio  $\approx \frac{\Delta R^{1\omega}_{xx}}{\Delta R^{1\omega}_{xy}}$. The second term of Eq. \ref{equation3} arises from the effect out of plane torque acting on the magnetization due to combined Oersted and field like effective field and the prefactor $\Delta R_{xx}^{1\omega}$ is the maximum change in the first harmonic resistance as the FM magnetization is flipped from along the current to perpendicular to current configuration.
\begin{figure}[h!]
\includegraphics[width=.5\textwidth]{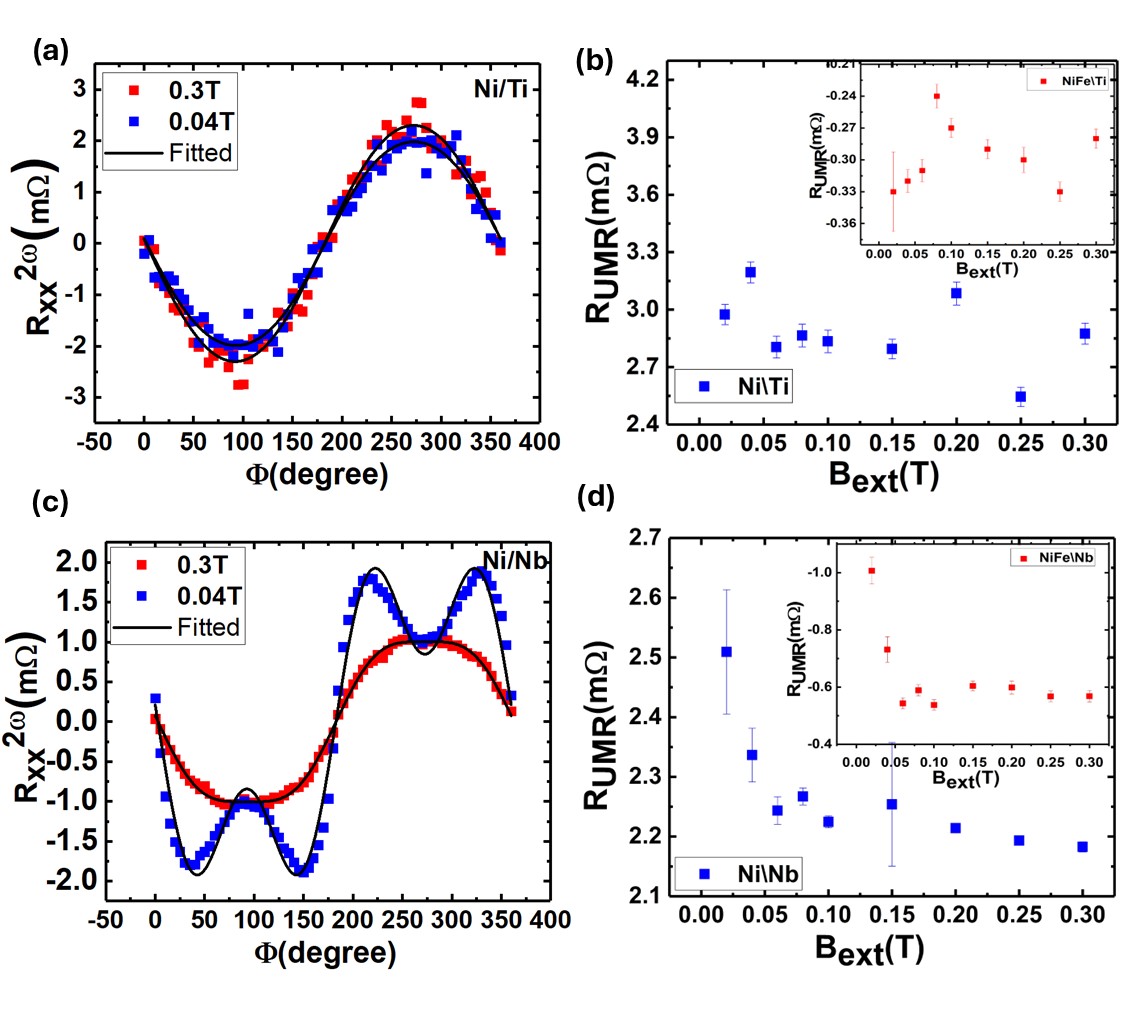}
    \caption{The second harmonic Hall Resistance($R^{2\omega}_{xx}$) as a function of  In Plane Rotation Of External Magnetic Field Of All devices. (a),(c) Angular variation of ($R^{2\omega}_{xx}$) at different external magnetic fields of Ni/Ti and Ni/Nb devices respectively. (b),(d)$R_{UMR}+R_{\nabla T}$ at different external magnetic fields of Ni/Ti and Ni/Nb devices.  }
    \label{4}
\end{figure}
Our measurements of $R^{2\omega}_{xx}$ on Ni/Ti and Ni/Nb devices for the two extreme magnetic fields are shown in Fig.\ref{4} (a) and (c) respectively,  and the corresponding fits to Eq.\ref{equation3} shown as solid lines, are found to be in good agreement. However we observe a striking feature in the low field behavior of Ni/Nb bilayer, that exhibits four extrema point which arises due to a pronounced  FL+Oe torque term, as compared to the Ni/Ti devices where the feature is absent.  However for both NiFe/Ti and NiFe/Nb, we observe the effects of FL+Oe torque ( see Supplementary ). As mentioned in the previous section,  FL torque is normally  dominated by interface properties of FM/NM like Rashba Edelstein effect, rather than from bulk property of the normal metal like OHE conductivity\cite{PhysRevB.92.064426, PhysRevApplied.13.054014}.  Our results suggests that the interface of Nb with both FMs( Ni and NiFe) and that of Ti with NiFe generates observable (FL+Oe) torque, whereas interface of Ti with Ni has negligible contribution from the effect.

The  values of $R_{UMR}$ extracted from the fits for different $B_{ext}$  are shown in Fig \ref{4} (b) and (d) for Ni/Ti and Ni/Nb respectively and the insets shows the corresponding results for NiFe/Ti and NiFe/Nb.
We observe a significant enhancement in the magnitude of $R_{UMR}$  in Ni/Ti and Ni/Nb bilayers compared to those of NiFe/Ti and NiFe/Nb. Previous report of UOMR on Cu/Co bilayer was explained on the basis of  scattering between OAM and the substantial spin polarization associated with the ferromagnetic magnetization in Co. We propose, that increased $R_{UMR}$ in bilayers with Ni in comparison to those of  NiFe, is arising from  orbital-to-spin conversion in Ni which consistent with the relatively large orbital torque as shown in previous section. Another important finding revealed in our experiments is that  Ni/Nb exhibits a pronounced low field enhancement in the values of $R_{UMR}$ , eventually reaching a saturation value at higher fields. This feature is not observed in Ni/Ti bilayers where the $R_{UMR}$ values doesnot show any systematic dependence on magnetic field. The corresponding devices with NiFe ( shown in insets) follow the same trends as that of Ni bilayers.

As discussed in the earlier report on UOMR\cite{PhysRevResearch.4.L032041}, a key difference between orbital and spin UMR is that low-field enhancement of UMR is not seen in LM/FM bilayers, unlike in HM/FM bilayers. This distinction arises because, in HM/FM systems, spin current can excite or annihilate magnons in the FM, leading to electron-magnon scattering that enhances UMR at low fields. In contrast, this mechanism is less active in LM/FM bilayers, resulting in a reduced low-field response. 
We speculate that the presence of the low field enhancement of $R_{UMR}$ for Nb as LM is indicative of the presence spin currents in addition to OAM current in Nb as compared to Ti where only orbital current is generated due to flow of applied charge current.
The absence of orbital-magnon coupling can be attributed to the nature of magnons as bosonic spin excitations. Since magnons represent collective spin oscillations, they do not directly couple to orbital angular momentum, leading to a decoupling between orbital effects and magnonic spin dynamics. However, an orbital current can directly influence the electrical conductivity in a ferromagnet through orbital-selective
s→d transitions. This interaction enhances the UMR due to increased scattering arising from orbital contributions, leading to the observed amplification in systems where orbital scattering plays a significant role. This orbital scattering effect, which enhances UMR, is absent in single ferromagnetic (FM) layer devices. We note that measurements carried out on single layer of Ni doesnot show any effect of UMR further corroborating the effect of orbital current generated in the LM layers. 
\begin{figure}[h!]
\includegraphics[width=.5\textwidth]{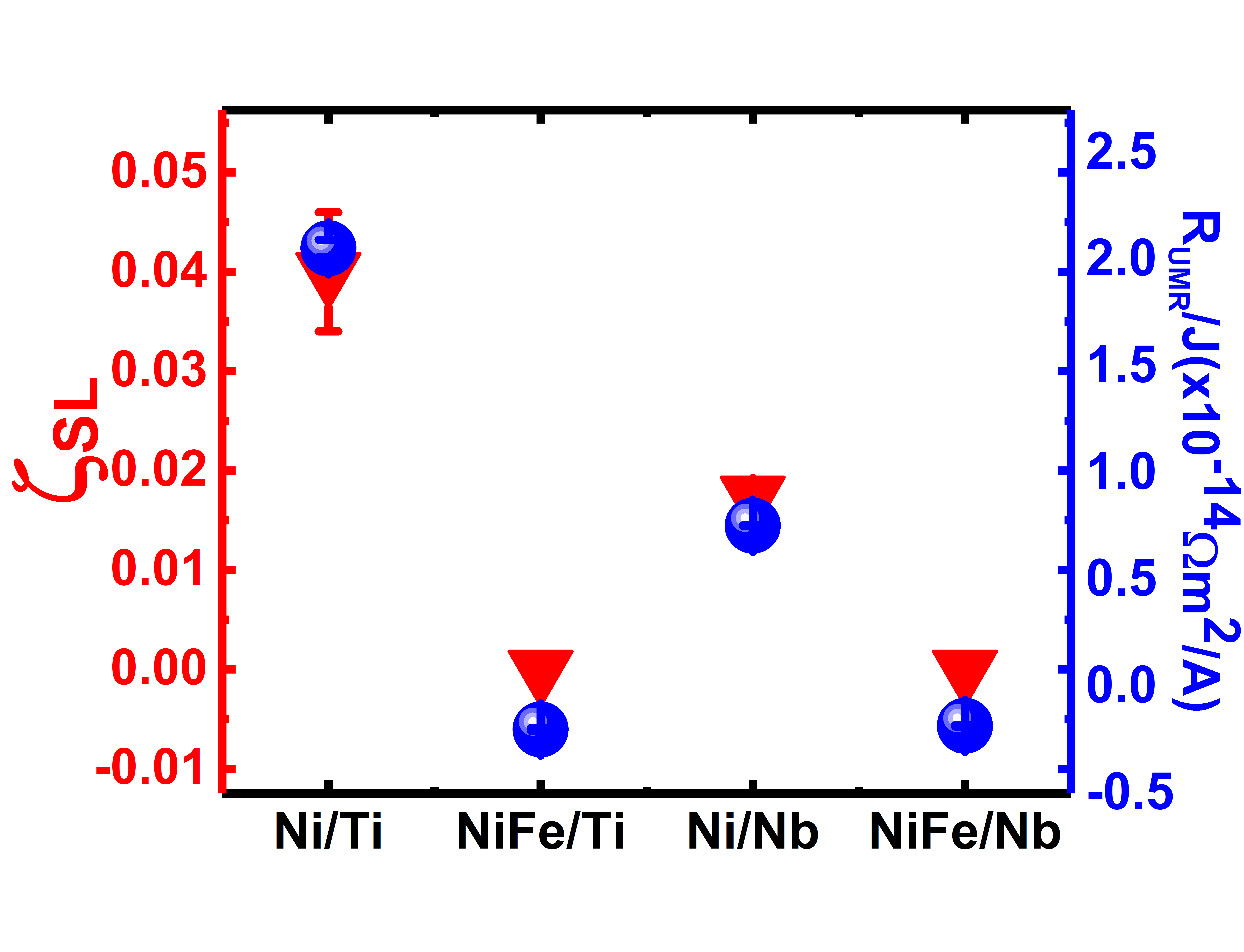}
    \caption{Correlation between Torque efficiency ($\zeta_{SL}$) (inverted triangles) and UMR per current density ($R_{UMR}/J$)(circles) for all devices. The values of $\zeta_{SL}$ are shown in Table S1 and values of $R_{UMR}$  are shown in Table S2 in the supplementary material, Notes 4 and 6, respectively.}
    \label{5}
\end{figure}
$R_{UMR}$ extracted from the $2\omega$ response is nonlinear in current response and also,  as low field magnon induced enhancement is present in some LM materials like Nb, for comparison  between different devices the high field values of $R_{UMR}$ normalized by current densities  should be considered.  Figure \ref{5} shows on the same plot both $\zeta_{DL}$ and $R_{UMR}/J_{LM}$ for all the devices,  which  reveals a prominent correlations between the $R_{xy}^{2\omega}$ and  $R_{UMR}/J_{LM}$ which is expected as both originates from OAM currents in the LM layers. In our experiments,  both quantities are higher for Ti device compared to that of Nb devices. Also, we observe that even though Orbital Torque was absent for the NiFe devices, a significantly small but finite and negative values of $R_{UMR}$ is observed. This can be understood from the fact that torque on the FM layer can be generated only when the orbital current is able to penetrate the FM layer with a reasonable efficiency so that it can transfer the component of angular momentum perpendicular to the FM magnetization.  The interface transparency and diffusion lengths play a decisive role. However, $R_{UMR}$ arises from the  accumulation of the OAM  and is not expected to be sensitive to interfaces.




To summarize, we report  experimental observation of OAM current originating in LM layers of Ti and Nb in Ni/LM from the measurement of second harmonic response of both longitudinal and transverse voltages. We observe a large Orbital Torque effect and UOMR as extracted from variation of transverse and longitudinal voltages respectively. 
The control samples of NiFe/LM bilayers and single layer Ni further establishes the role of LM in generating the OAM current through OHE. Nb samples exhibits and additional feature of  magnon induced enhancement in resistance  was not observed in Ti samples, although the absolute values of both the orbital torque efficiency and Unidirectional Magnetoresistance is higher in Ti. This observations will excite further interests for  experiments with wide range of LM layers along with the development of a theoretical framework for orbital and spin angular momentum interaction.
\\\\\\

\section*{Supplementary Material}
See the supplementary material for further details on the experimental aspects of this work. This includes a description of the measurement techniques, anomalous and planar Hall effect (AHE and PHE) analysis, structural characterization, as well as tabulated values of effective fields, thermal contributions, and unidirectional magnetoresistance data.\\\\\\

The authors thank IISER Kolkata, an autonomous research and teaching institution funded
by the MoE, Government of India for providing the financial support and infrastructure. The authors also thank CSIR and UGC for providing fellowship.

\nocite{*}
\bibliography{aipsamp}

\end{document}


\preprint{}
\begin{center} 
\Huge{Supplementary Information}
\end{center}
\title{Evidence of orbital Hall current induced correlation in second harmonic response of longitudinal and transverse voltage in light metal-ferromagnet bilayers.  }
\author{Dhananjaya Mahapatra}
 \altaffiliation{Department of Physical Sciences,
 Indian Institute of Science Education and Research Kolkata, Mohanpur, West Bengal,741246, INDIA.}
 \author{Harekrishna Bhunia}
 \altaffiliation{Department of Physical Sciences,
 Indian Institute of Science Education and Research Kolkata, Mohanpur, West Bengal,741246, INDIA.}
\author{Abu Bakkar Miah}
 \altaffiliation{Department of Physical Sciences,
 Indian Institute of Science Education and Research Kolkata, Mohanpur, West Bengal,741246, INDIA.}
 \author{Soumik Aon}
 \altaffiliation{Department of Physical Sciences,
 Indian Institute of Science Education and Research Kolkata, Mohanpur, West Bengal,741246, INDIA.}
\author{Partha Mitra}
 \email{pmitra@iiserkol.ac.in}
\altaffiliation{Department of Physical Sciences,
 Indian Institute of Science Education and Research Kolkata, Mohanpur, West Bengal,741246, INDIA.}

\date{\today}

\maketitle
\section*{Supplementary Note 1: Experimental technique.}

\begin{figure}[h!]
    \includegraphics[width=1.1\textwidth]{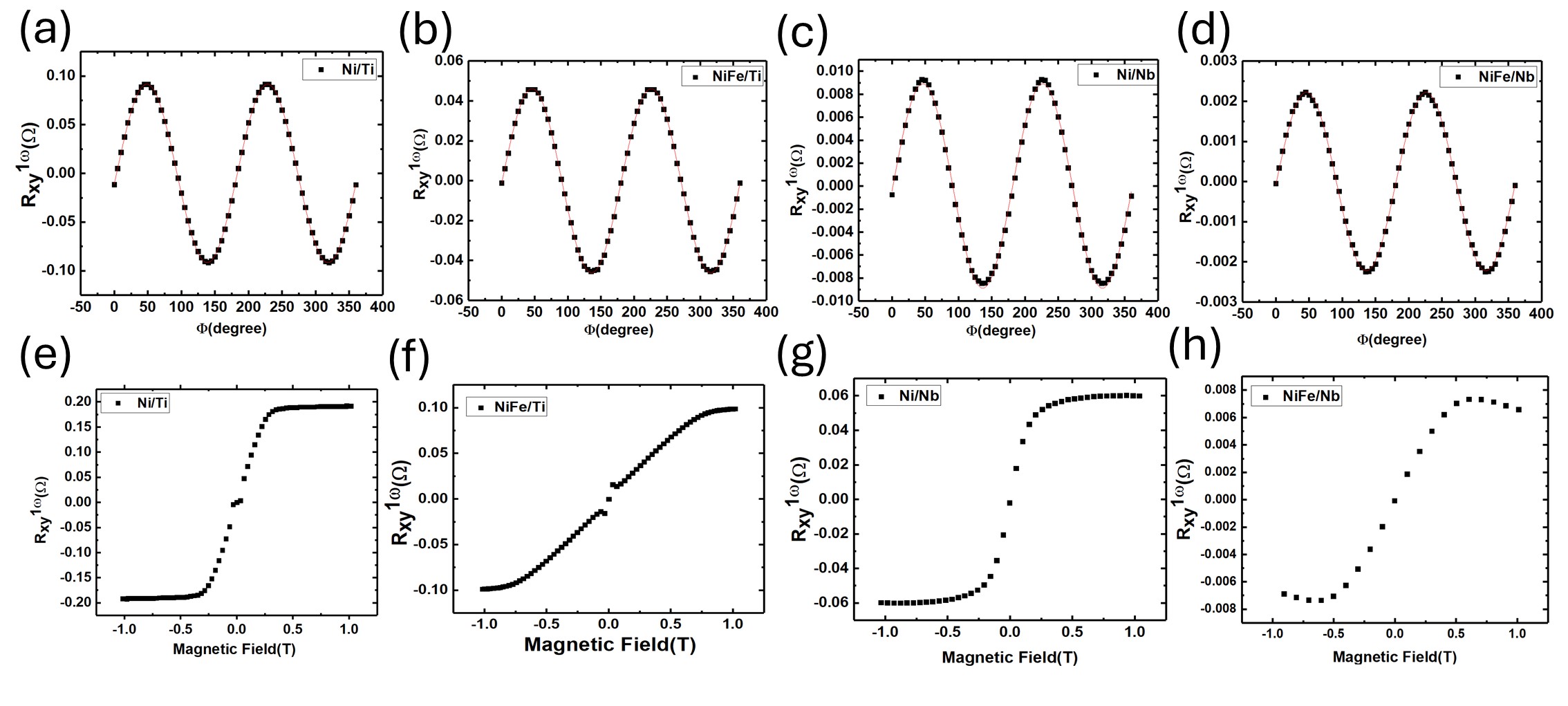}
    \caption{a,b,c,d In plane Angular variation of $R_{xy}^{1\omega}$ of Ni/Ti, NiFe/Ti, Ni/Nb, NiFe/Nb devices. e,f,g,h Out of plane magnetic field scan of Ni/Ti, NiFe/Ti, Ni/Nb, NiFe/Nb devices.}
    \label{1}
\end{figure}

All FM/LM devices are patterned in the shape of standard Hall bars(Fig. \ref{1}(c)) with current channel dimensions of $(2\times 30) \mu m^2$ and voltage pickup line widths of $0.5 \mu$ using ebeam lithography ( ELPHY Quantum, Raith ) on Si/SiO2(300nm) substrates with all FM and LM layers deposited for a thickness of 10nm. The FM layer of either Ni or NiFe was first deposited by the thermal evaporation technique at the rate of (0.8-0.9){\AA} /s. Thereafter, the LM layer of either Ti or Nb was deposited on top of the FM by DC sputtering at rates of 0.5-0.6\AA /s. Ar ion milling was performed before all deposition to clean the interface and promote good Ohmic contact between FM and LM. Finally, contact pads were patterned on the Hall bar electrodes using UV photolithography and Cr/Au was deposited using ebeam evaporation. X-ray diffraction (XRD) with a Cu-$K_\alpha$(1.54\AA) source confirms fcc structure for Ti and bcc for Nb. The bare resistivities of each metallic layers were determined using  Van der Pauw technique  and found to be $\rho_{Ti}=80,\rho_{Nb}=35,\rho_{Ni}=130,\rho_{NiFe}=170 \mu\Omega.cm$. 
For harmonic measurements a constant a.c. current $I(t)=I_o sin(2\omega t)$ is sourced through the bilayer  Hall bar with $\omega=2\pi X 13$rad/s using a Keithley 6221A AC-DC current source and the amplitudes of first and second harmonic voltage responses are recorded across both longitudinal ($V_{xx}^\omega, V_{xy}^{2\omega}$) and transverse ($V_{xy}^\omega,V_{xy}^{2\omega}$) voltage leads of the Hall bars using a SR830 Lockin amplifier (Fig. \ref{1}c). The corresponding resistances are defined as the ratio of the voltage amplitude to current amplitude, in particular $R_{xx}^{2\omega}=V_{xx}^{2\omega}/I_o$ and $R_{xy}^{2\omega}=V_{xy}^{2\omega}/I_o$. As the FM and NM layers are in ohmic contact, the applied current is distributed between the two layers depending on the resistivity of the layers and the current density in each layers can be estimated as, 
$J_{LM}$=$\frac{I_{o}}{(wt)[1+(\frac{\rho_{LM}}{\rho_{FM}})]}$ and $J_{FM}$=$\frac{I_{o}}{(wt)[1+(\frac{\rho_{FM}}{\rho_{LM}})]}$. 
The value of $I_o$ is suitably chosen so that the estimated $J_{LM}$ in devices with same LM layer but different FM layers (Ni or NiFe) remains same, ensuring same orbital Hall current. Typical current densities for our experiments are in the range of $10^{11} A/m^2$. Our goal is to investigate  the harmonic  responses of the FM/LM bilayers as the FM magnetisation is rotated with respect to the polarization of the OAM current arising from OHE in the LM layer.
The devices are mounted on a homemade setup with 2D vector electromagnet ( DEXING, China) with the Hall bar lying in the plane of the magnetic field vector and the current channel along the $x$-component of magnetic field. The measurement of Planar Hall effect confirms that the magnetisation of the FM is indeed rotating with the magnetic field ( See Supplementary note 2). A typical variation of the second harmonic resistances for a Ni/Ti device from $B=0.3$T  shown in Fig. \ref{1}(d) reveals that the variation of $R_{xy}^{2\omega}$ is opposite to that of $R_{xx}^{2\omega}$. As $R_{xy}^{2\omega}$ bears the effect of orbital torque transferred to the perpendicular component of magnetisation from the OAM current, it reaches maximum value  at $\Phi=90^o, 270^o$. $R_{xx}^{2\omega}$ on the other hand arises due to GMR like effect where the resistance depends on the relative orientation of OMR polarization and majority spins of the FM and and reaches extrema for $\Phi=0^o and 180^o$. In the subsequent section we present through analysis of the second harmonic data.

\newpage
\section*{Supplementary Note 2: Anomalous and planar Hall effect of FM/light metal bilayers.}
We measured the 1st harmonic planar hall effect under a constant external 0.3T magnetic field in order to achieve the ideal alignment of current and magnetization. Then, in order to analyze spin orbit torques (SOT) by in plane harmonic hall measurements(eq1 in the main text), we measure the first harmonic anomalous hall resistance($R_{AHE}^{1\omega}$).Supplementary Figures 2a-d show azimuthal angle-dependent $1^{st}$ harmonic hall resistance $R^{1\omega}_{xy}$of the FM(Ni and NiFe)/LM(Ti and Nb) samples, In order to characterize the ferromagnetic sample and achieve perfect alignment between the current and the external magnetic field, it is measured under a constant external magnetic field of 0.3T fitted by $R^{1\omega}_{xy}=R_{PHE}sin(2\Phi)$. Supplementary Figure 2e-h,shows $1^{st}$ harmonic resistance $R^{1\omega}_{xy}$ as a function of perpendicular external magnetic field$(B_z)$.The intersection between low field straight line fit and high field straight line fit gives the value $B^k_{eff}$ and $R_{AHE}$\cite{10.1063/1.5027855}.
\newpage
\section*{Supplementary Note 3: XRD pattern of Ti and Nb thinfilm.}
\begin{figure}[h!]
    \includegraphics[width=1.1\textwidth]{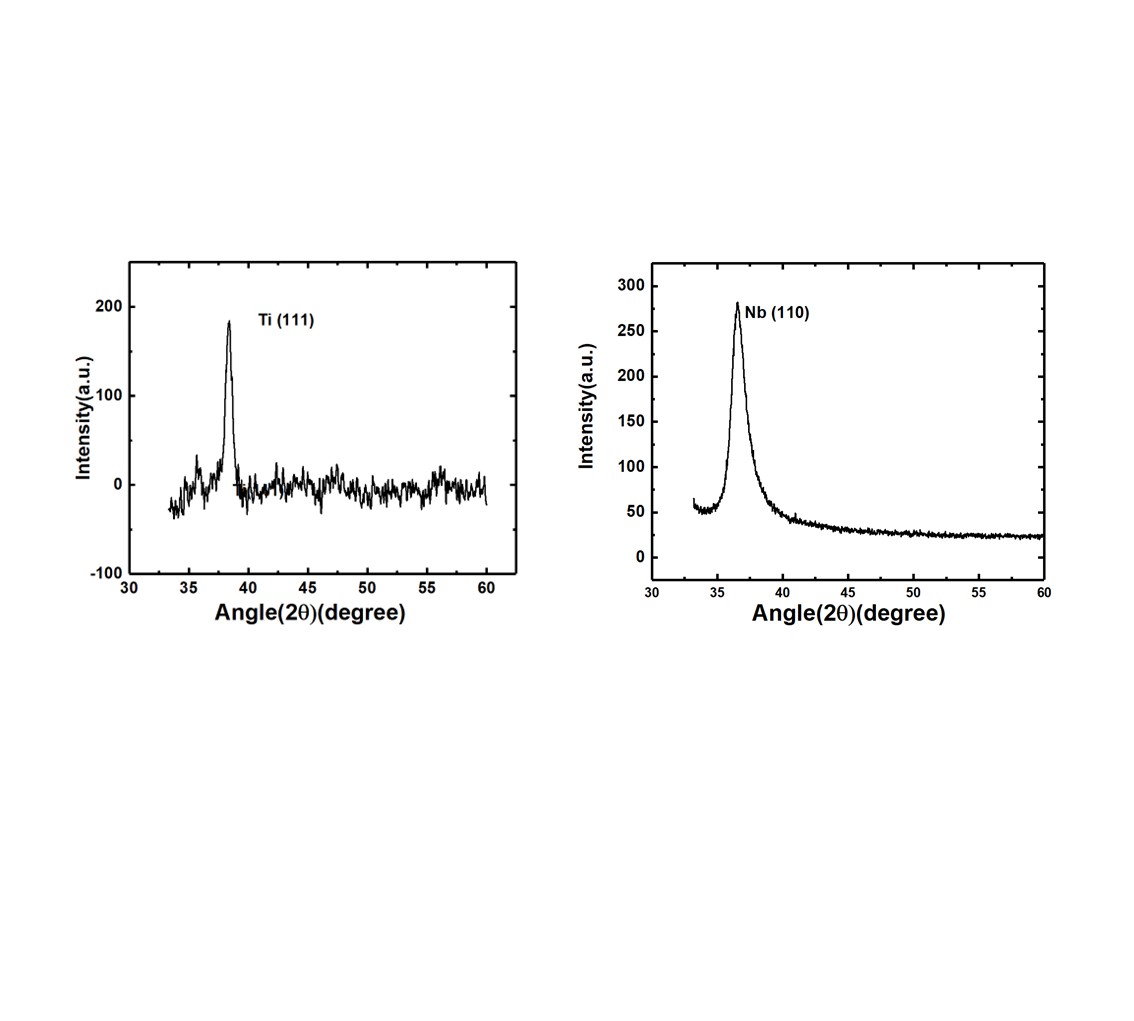}
    \caption{X-ray diffraction (XRD) patterns of Ti and Nb thin films. The sharp peaks observed at $2\theta = 38.3$ and $36.4$ correspond to the (111) planes of face-centered cubic (FCC) Ti and (110) planes of body-centered cubic (BCC) Nb respectively.}
    \label{2}
\end{figure}

\newpage
\section*{Supplementary Note 4:SL effective field with all heating term value. }

The angular variations of $R^{2\omega}_{xy}$ in the inplane magnetic field are entirely explained by equation.  $ R_{xy}^{2\omega}(\Phi)=[R_{AHE}(\frac{B_{SL}}{B_{ext}+B_{eff}^k})+\alpha B_{ext} + R_{\nabla T}]cos(\Phi) +2R_{PHE}(\frac{B_{FL}+B_{Oe}}{B_{ext}})cos(2\Phi)cos(\Phi)$\cite{PhysRevB.99.195103, JOUR, AHN202312}. By extracting the cosine contribution from the signal and subsequently visualizing all cosine contributions against varying external magnetic fields.Finally, we extract $B_{SL}$, $\frac{\alpha}{R_{AHE}}$ and $\frac{R_{\Delta T}}{R_{AHE}}$ which is shown in TABLE\ref{demo-table}. Where $B_{SL}$ is SL effective field, $\alpha$ is the Ordinary Nernst(ONE) co-efficient, and $R_{\nabla T}$ is the anomalous Nernst(ANE) co-efficient.$\alpha$ and $R_{\nabla T}$ are coming from heating effect.

\begin{widetext}
\begin{table}[!h]
\begin{center}
\caption{}
\label{demo-table}
\begin{tabular}{||p{2cm}|p{3.5cm}|p{3.5cm}|p{3.5cm}|p{3.5cm}||} 
 \hline
    DEVICES   & $B_{SL}(\mu T)$   & $\frac{\alpha}{R_{AHE}}$ & $\frac{R_{\nabla T}}{R_{AHE}}$ & $\zeta_{SL}$\\ [2 ex] 
 \hline\hline
  Ni & 3$\pm$1  & -6.10225$\times 10^{-5}$ $\pm$3.613384$\times 10^{-5}$ & -0.00102 $\pm$1.84391$\times 10^{-5}$ & 0.0006$\pm$0.0001 \\ 
 \hline
  NiFe & 5$\pm$27  & -3.57808$\times 10^{-5}$ $\pm$7.12626$\times 10^{-5}$ & -0.00161 $\pm$5.86704$\times 10^{-5}$ &   $\approx 0$  \\ 
 \hline
 Ni/Ti & 100 $\pm$ 15   & -1.15843$\times 10^{-4}$ $\pm$4.77366$\times 10^{-4}$ & -0.00691 $\pm$2.26216$\times 10^{-4}$ & 0.04 $\pm$ 0.006 \\
 \hline
 Ni/Nb & 1900 $\pm$ 130  & -0.00668 $\pm$0.00109 & -0.01474 $\pm$5.74833$\times 10^{-4}$ & 0.018$\pm$0.001  \\
 \hline
 NiFe/Ti & -481$\pm$973 & -7.24799$\times 10^{-4}$ $\pm$0.00126 & -0.00488 $\pm$0.00135 &   $\approx 0$ \\
 \hline
 NiFe/Nb & -1320$\pm$1530 & -0.00349$\pm$0.00501 & -0.00802 $\pm$0.00315 &  $\approx 0$ \\
 \hline
  Ni/Pt & 23$\pm$2 & -3.30738$\times 10^{-5}$$\pm$ 3.8728$\times 10^{-5}$ & -0.00165$\pm$2.35284 $\times 10^{-5}$&  0.03$\pm$0.001 \\
 \hline
 NiFe/Pt & 700$\pm$200 & -0.00139$\times 10^{-4}$$\pm$ 3.8728$\times 10^{-5}$ & -0.00218$\pm$6.3012 $\times 10^{-4}$&  0.13$\pm$0.03 \\
 \hline
\end{tabular}
\end{center}
\end{table}
\end{widetext}


\newpage
\section*{Supplementary Note 5:Field-like Torques of the FM/LM Bilayers. }

\begin{figure}[h!]
    \includegraphics[width=1\textwidth]{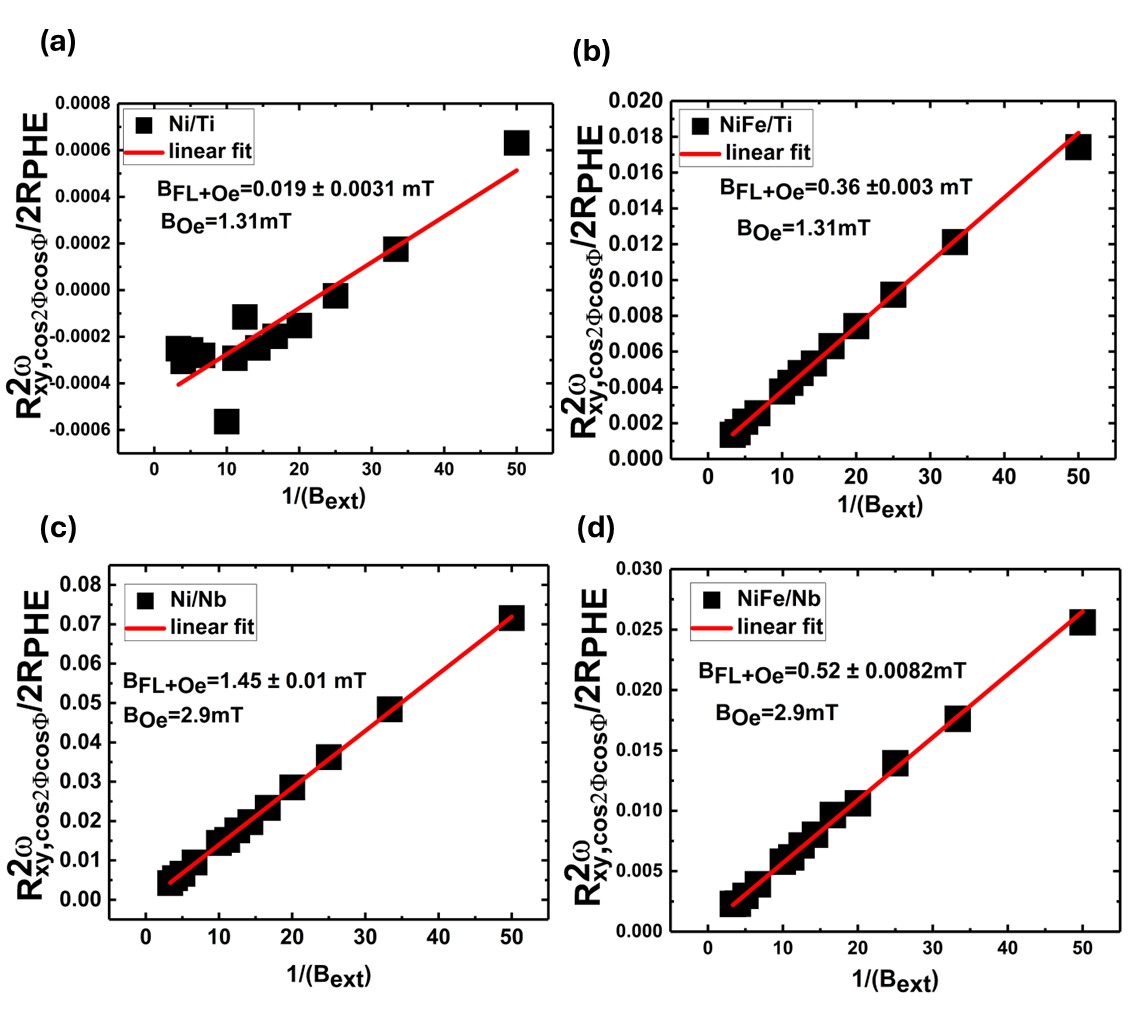}
    \caption{ }
    \label{S6}
\end{figure}

Supplementary Figure(\ref{S6})Shows $cos(2\Phi)cos(\Phi)/2R_{phe}$ for a. Ni/Nb, b.Ni/Ti, C. NiFe/Nb, and d. NiFe/Ti samples as a function of $1/B_{ext}$. The slope of the linear fit provides $B_{FL+Oe}$. We calculated $B_{Oe}$ using $B_{Oe}$ =$\frac{\mu_0 I_{NM}}{2w}$, where w is the width of the hall bar devices, and found that $B_{Oe}$ is the dominant contribution across all devices.

\newpage
\section*{Supplementary Note 6:$R_{UMR}$ value with heating term. }
The angular variations of $R^{2\omega}_{xx}$ in the inplane magnetic field are entirely explained by equation. $ R_{xx}^{2\omega}=R^* sin(\Phi)-2\Delta R_{xx}^{1\omega} \frac{B_{FL}+B_{Oe}}{B_{ext}} cos^2(\Phi) sin(\Phi)$. The field-dependent variations of $R_{xx}^{2\omega}$were fitted using the proposed model equation, enabling the extraction of the parameter $R^*$, as listed in Table(\ref{table2}). Subsequently, the unidirectional magnetoresistance ($R_{UMR}$) was calculated using the relation $R_{UMR}=R^*-gR_{\nabla T}$ , where g denotes the geometric factor, approximately equal to 4. The final values of $R_{UMR}$are summarized in Table(\ref{table2}).

\begin{widetext}
\begin{table}[!h]
\begin{center}
\caption{}
\label{table2}
\begin{tabular}{||p{2cm}|p{3.5cm}|p{3.5cm}|p{3.5cm}||} 
 \hline
    DEVICES   & $R^*(m\Omega)$   & $g R_{\nabla T}(m\Omega)$ & $R_{UMR}(m\Omega)$ \\ [2 ex] 
 \hline\hline
  Ni & -0.77 &  -0.342 & -0.428$\pm$0.007 \\ 
  \hline
  Ni/Ti & -2.3 &  -5.17 & 2.87$\pm$0.05 \\ 
  \hline
  Ni/Nb & -1&  -3.18 & 2.18$\pm$ 0.008 \\
  \hline
   NiFe/Ti & -2.15&  -1.87 & -0.28$\pm$ 0.01 \\
  \hline
   NiFe/Nb & -0.83 &  -0.26 & -0.57$\pm$0.01 \\
  \hline
\end{tabular}
\end{center}
\end{table}
\end{widetext}

\newpage
\section*{Supplementary Note 7: Angular dependence of $R_{xx}^{2\omega}$ of Py/Ti and Py/Nb devices. }
The angular variations of $R^{2\omega}_{xx}$ in the inplane magnetic field of NiFe/Ti and NiFe/Nb are fitted by equation $ R_{xx}^{2\omega}=R^* sin(\Phi)-2\Delta R_{xx}^{1\omega} \frac{B_{FL}+B_{Oe}}{B_{ext}} cos^2(\Phi) sin(\Phi)$.
\begin{figure}[h!]
    \includegraphics[width=1\textwidth]{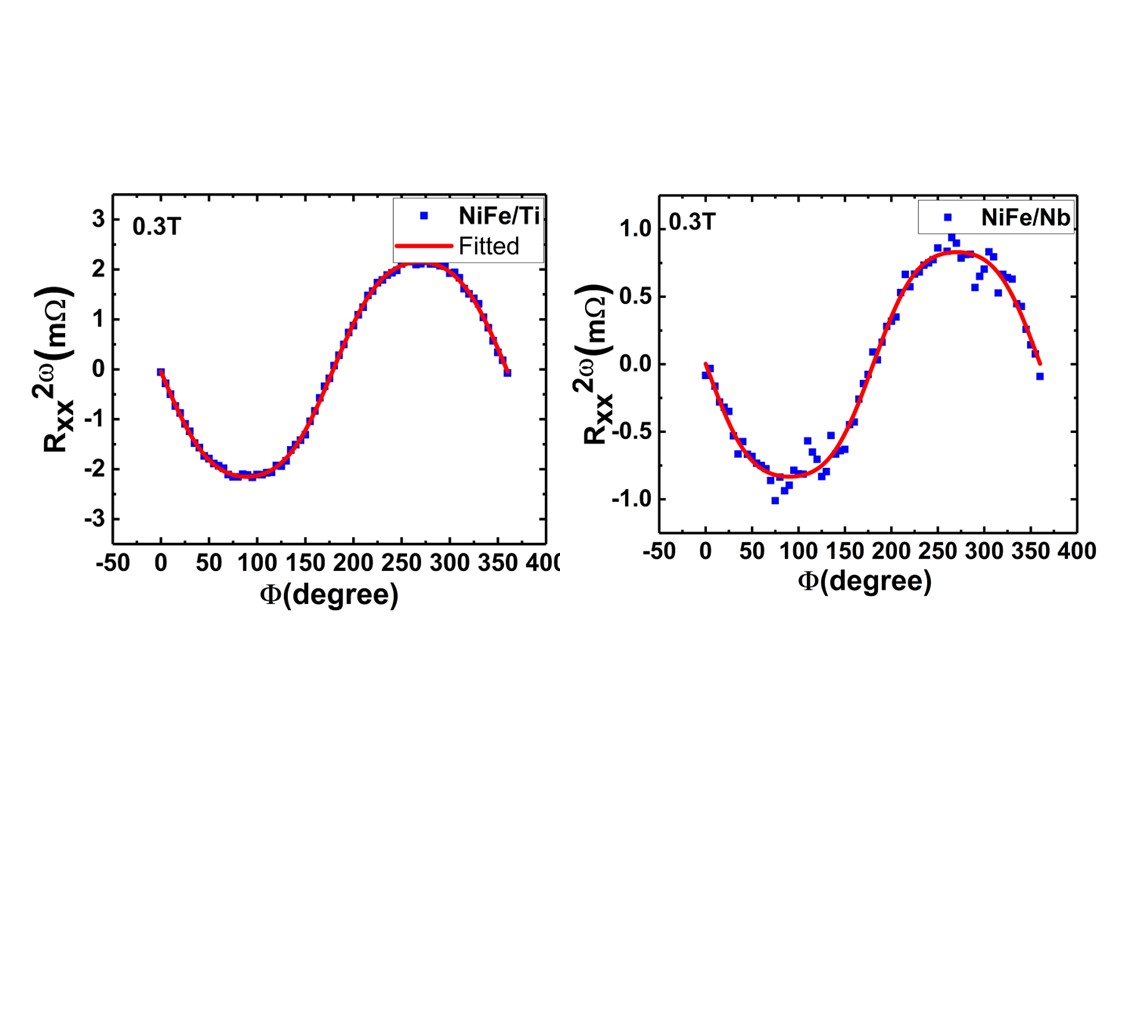}
    \caption{ }
    \label{S7}
\end{figure}

\newpage
\nocite{*}

\bibliography{aipsupl}